\documentclass[sigconf,letterpaper,screen,balance=True,authorversion]{acmart}

\usepackage[utf8]{inputenc} 

\acmYear{2019}\copyrightyear{2019}
\setcopyright{acmcopyright}
\acmConference[CoNEXT '19]{CoNEXT '19: International Conference On Emerging Networking Experiments And Technologies}{December 9--12, 2019}{Orlando, FL, USA}
\acmBooktitle{CoNEXT '19: International Conference On Emerging Networking Experiments And Technologies, December 9--12, 2019, Orlando, FL, USA}
\acmPrice{15.00}
\acmDOI{10.1145/3359989.3365416}
\acmISBN{978-1-4503-6998-5/19/12}

\PassOptionsToPackage{spaces,hyphens}{url}
\usepackage[inline]{enumitem}

\usepackage{booktabs}
\usepackage{multirow}
\usepackage{units}
\usepackage{lipsum}

\newcommand{\eg}{e.g.,}

\newcommand{\ie}{i.e.,}

\newcommand{\sref}[1]{Section~\ref{#1}} 
\newcommand{\fig}[1]{Figure~\ref{#1}}
\newcommand{\tab}[1]{Table~\ref{#1}} 
\newcommand{\afblock}[1]{\noindent{\textbf{#1}}} 
\newcommand{\takeaway}[1]{\noindent{\textbf{Takeaway.}} \textit{#1}}

\providecommand\unit[2]{#1\,#2}
\begin{document}
\def\ABDurationMeanNet{5.94} %
\def\ABDurationMedianNet{5.28} %
\def\ABDurationMeanWork{8.67} %
\def\ABDurationMedianWork{7.84} %
\def\ABDurationMeanLab{9.95} %
\def\ABDurationMedianLab{8.84} %
\def\ABLabConditions{83} %
\def\ABMinVotesLab{10} %
\def\ABMinVotesNet{0} %
\def\ABMinVotesWork{4} %
\def\ABMinVotesAll{12} %
\def\ABAvgVotesLab{11.8} %
\def\ABAvgVotesNet{3.7} %
\def\ABAvgVotesWork{10.5} %
\def\ABAvgVotesAll{15.9} %
\def\ABVotesLab{980} %
\def\ABVotesNet{2170} %
\def\ABVotesWork{6058} %
\def\ABRelVotesLab{10.6} %
\def\ABRelVotesNet{23.6} %
\def\ABRelVotesWork{65.8} %
\def\ABFilterLab{35} %
\def\ABFilterLabi{35} %
\def\ABFilterLabii{35} %
\def\ABFilterLabiii{35} %
\def\ABFilterLabiv{35} %
\def\ABFilterLabv{35} %
\def\ABFilterLabvi{35} %
\def\ABFilterLabvii{35} %
\def\ABFilterNet{218} %
\def\ABFilterNeti{217} %
\def\ABFilterNetii{210} %
\def\ABFilterNetiii{196} %
\def\ABFilterNetiv{171} %
\def\ABFilterNetv{170} %
\def\ABFilterNetvi{159} %
\def\ABFilterNetvii{155} %
\def\ABFilterWork{487} %
\def\ABFilterWorki{471} %
\def\ABFilterWorkii{441} %
\def\ABFilterWorkiii{355} %
\def\ABFilterWorkiv{268} %
\def\ABFilterWorkv{268} %
\def\ABFilterWorkvi{239} %
\def\ABFilterWorkvii{233} %
\def\ABGenderMaleLab{80.0} %
\def\ABGenderFemaleLab{20.0} %
\def\ABGenderDiversLab{0.0} %
\def\ABGenderNoInformationLab{0.0} %
\def\ABExpertiseBeginnerLab{0.0} %
\def\ABExpertiseNormalLab{31.4} %
\def\ABExpertiseExpertLab{68.6} %
\def\ABAgeLeveliLab{60.0} %
\def\ABAgeLeveliiLab{40.0} %
\def\ABAgeLeveliiiLab{0.0} %
\def\ABHoursLeveliLab{2.9} %
\def\ABHoursLeveliiLab{31.4} %
\def\ABHoursLeveliiiLab{65.7} %
\def\ABGenderMaleNet{76.1} %
\def\ABGenderFemaleNet{20.0} %
\def\ABGenderDiversNet{0.0} %
\def\ABGenderNoInformationNet{3.9} %
\def\ABExpertiseBeginnerNet{2.6} %
\def\ABExpertiseNormalNet{36.1} %
\def\ABExpertiseExpertNet{61.3} %
\def\ABAgeLeveliNet{58.1} %
\def\ABAgeLeveliiNet{34.2} %
\def\ABAgeLeveliiiNet{7.7} %
\def\ABHoursLeveliNet{7.1} %
\def\ABHoursLeveliiNet{41.3} %
\def\ABHoursLeveliiiNet{51.6} %
\def\ABGenderMaleWork{78.1} %
\def\ABGenderFemaleWork{21.9} %
\def\ABGenderDiversWork{0.0} %
\def\ABGenderNoInformationWork{0.0} %
\def\ABExpertiseBeginnerWork{2.6} %
\def\ABExpertiseNormalWork{58.8} %
\def\ABExpertiseExpertWork{38.6} %
\def\ABAgeLeveliWork{26.2} %
\def\ABAgeLeveliiWork{66.5} %
\def\ABAgeLeveliiiWork{7.3} %
\def\ABHoursLeveliWork{1.7} %
\def\ABHoursLeveliiWork{43.8} %
\def\ABHoursLeveliiiWork{54.5} %
\def\MeanVerdictDslTcpTcpp{0.21} %
\def\MeanVerdictLteTcpTcpp{0.59} %
\def\MeanVerdictGggTcpTcpp{0.45} %
\def\MeanVerdictDaagcTcpTcpp{-0.16} %
\def\MeanVerdictMssTcpTcpp{0.45} %
\def\MeanVerdictDslTcpQuic{0.61} %
\def\MeanVerdictLteTcpQuic{0.88} %
\def\MeanVerdictGggTcpQuic{0.69} %
\def\MeanVerdictDaagcTcpQuic{0.53} %
\def\MeanVerdictMssTcpQuic{0.98} %
\def\MeanVerdictDslTcppQuic{0.38} %
\def\MeanVerdictLteTcppQuic{0.72} %
\def\MeanVerdictGggTcppQuic{0.62} %
\def\MeanVerdictDaagcTcppQuic{0.55} %
\def\MeanVerdictMssTcppQuic{0.92} %
\def\MeanVerdictDslTcpbbrQuicbbr{0.39} %
\def\MeanVerdictLteTcpbbrQuicbbr{0.68} %
\def\MeanVerdictGggTcpbbrQuicbbr{0.57} %
\def\MeanVerdictDaagcTcpbbrQuicbbr{0.77} %
\def\MeanVerdictMssTcpbbrQuicbbr{0.80} %
\def\ABLabNetDiffMean{0.23} %
\def\ABLabNetDiffMedian{0.22} %
\def\ABLabWorkDiffMean{0.20} %
\def\ABLabWorkDiffMedian{0.28} %
\def\RateDurationMeanNet{7.32} %
\def\RateDurationMedianNet{6.81} %
\def\RateDurationMeanWork{10.68} %
\def\RateDurationMedianWork{9.68} %
\def\RateDurationMeanLab{12.41} %
\def\RateDurationMedianLab{11.92} %
\def\RateLabConditions{182} %
\def\RateMinVotesWorkNet{0} %
\def\RateAvgVotesWorkNet{1.7} %
\def\RateMinVotesFreetimeNet{0} %
\def\RateAvgVotesFreetimeNet{1.6} %
\def\RateMinVotesPlaneNet{0} %
\def\RateAvgVotesPlaneNet{1.9} %
\def\RateMinVotesWorkWork{9} %
\def\RateAvgVotesWorkWork{13.2} %
\def\RateMinVotesFreetimeWork{8} %
\def\RateAvgVotesFreetimeWork{13.0} %
\def\RateMinVotesPlaneWork{10} %
\def\RateAvgVotesPlaneWork{14.2} %
\def\RateMinVotesWorkLab{5} %
\def\RateAvgVotesWorkLab{5.5} %
\def\RateMinVotesFreetimeLab{4} %
\def\RateAvgVotesFreetimeLab{5.2} %
\def\RateMinVotesPlaneLab{5} %
\def\RateAvgVotesPlaneLab{5.5} %
\def\RateMinVotesWorkAll{14} %
\def\RateAvgVotesWorkAll{15.7} %
\def\RateMinVotesFreetimeAll{11} %
\def\RateAvgVotesFreetimeAll{15.3} %
\def\RateMinVotesPlaneAll{15} %
\def\RateAvgVotesPlaneAll{16.9} %
\def\RateVotesLab{980} %
\def\RateVotesNet{2208} %
\def\RateVotesWork{17192} %
\def\RateRelVotesLab{4.8} %
\def\RateRelVotesNet{10.8} %
\def\RateRelVotesWork{84.4} %
\def\RateFilterLab{35} %
\def\RateFilterLabi{35} %
\def\RateFilterLabii{35} %
\def\RateFilterLabiii{35} %
\def\RateFilterLabiv{35} %
\def\RateFilterLabv{35} %
\def\RateFilterLabvi{35} %
\def\RateFilterLabvii{35} %
\def\RateFilterNet{209} %
\def\RateFilterNeti{204} %
\def\RateFilterNetii{194} %
\def\RateFilterNetiii{172} %
\def\RateFilterNetiv{152} %
\def\RateFilterNetv{151} %
\def\RateFilterNetvi{140} %
\def\RateFilterNetvii{138} %
\def\RateFilterWork{1563} %
\def\RateFilterWorki{1494} %
\def\RateFilterWorkii{1321} %
\def\RateFilterWorkiii{1034} %
\def\RateFilterWorkiv{733} %
\def\RateFilterWorkv{723} %
\def\RateFilterWorkvi{661} %
\def\RateFilterWorkvii{614} %
\def\RateGenderMaleLab{80.0} %
\def\RateGenderFemaleLab{20.0} %
\def\RateGenderDiversLab{0.0} %
\def\RateGenderNoInformationLab{0.0} %
\def\RateExpertiseBeginnerLab{0.0} %
\def\RateExpertiseNormalLab{31.4} %
\def\RateExpertiseExpertLab{68.6} %
\def\RateAgeLeveliLab{60.0} %
\def\RateAgeLeveliiLab{40.0} %
\def\RateAgeLeveliiiLab{0.0} %
\def\RateHoursLeveliLab{2.9} %
\def\RateHoursLeveliiLab{28.6} %
\def\RateHoursLeveliiiLab{68.6} %
\def\RateGenderMaleNet{75.4} %
\def\RateGenderFemaleNet{18.8} %
\def\RateGenderDiversNet{0.7} %
\def\RateGenderNoInformationNet{5.1} %
\def\RateExpertiseBeginnerNet{2.2} %
\def\RateExpertiseNormalNet{36.2} %
\def\RateExpertiseExpertNet{61.6} %
\def\RateAgeLeveliNet{59.4} %
\def\RateAgeLeveliiNet{31.2} %
\def\RateAgeLeveliiiNet{9.4} %
\def\RateHoursLeveliNet{7.2} %
\def\RateHoursLeveliiNet{44.9} %
\def\RateHoursLeveliiiNet{47.8} %
\def\RateGenderMaleWork{67.8} %
\def\RateGenderFemaleWork{32.1} %
\def\RateGenderDiversWork{0.2} %
\def\RateGenderNoInformationWork{0.0} %
\def\RateExpertiseBeginnerWork{2.8} %
\def\RateExpertiseNormalWork{64.0} %
\def\RateExpertiseExpertWork{33.2} %
\def\RateAgeLeveliWork{28.5} %
\def\RateAgeLeveliiWork{61.9} %
\def\RateAgeLeveliiiWork{9.6} %
\def\RateHoursLeveliWork{4.4} %
\def\RateHoursLeveliiWork{47.4} %
\def\RateHoursLeveliiiWork{48.2} %
\def\RateLabNetDiffMean{5.28} %
\def\RateLabNetDiffMedian{5.68} %
\def\RateLabWorkDiffMean{4.77} %
\def\RateLabWorkDiffMedian{5.57} %
\def\MeanVotesWorkDslTcp{53.05} %
\def\MeanVotesWorkLteTcp{50.38} %
\def\MeanVotesWorkGggTcp{44.21} %
\def\MeanVotesWorkDslTcpp{53.74} %
\def\MeanVotesWorkLteTcpp{50.23} %
\def\MeanVotesWorkGggTcpp{45.13} %
\def\MeanVotesWorkDslTcpbbr{53.51} %
\def\MeanVotesWorkLteTcpbbr{49.61} %
\def\MeanVotesWorkGggTcpbbr{44.67} %
\def\MeanVotesWorkDslQuic{54.42} %
\def\MeanVotesWorkLteQuic{50.94} %
\def\MeanVotesWorkGggQuic{45.56} %
\def\MeanVotesWorkDslQuicbbr{54.31} %
\def\MeanVotesWorkLteQuicbbr{51.21} %
\def\MeanVotesWorkGggQuicbbr{45.58} %
\def\MeanVotesFreetimeDslTcp{54.62} %
\def\MeanVotesFreetimeLteTcp{49.62} %
\def\MeanVotesFreetimeGggTcp{45.99} %
\def\MeanVotesFreetimeDslTcpp{55.10} %
\def\MeanVotesFreetimeLteTcpp{49.30} %
\def\MeanVotesFreetimeGggTcpp{45.50} %
\def\MeanVotesFreetimeDslTcpbbr{55.11} %
\def\MeanVotesFreetimeLteTcpbbr{49.88} %
\def\MeanVotesFreetimeGggTcpbbr{45.80} %
\def\MeanVotesFreetimeDslQuic{55.93} %
\def\MeanVotesFreetimeLteQuic{51.62} %
\def\MeanVotesFreetimeGggQuic{47.83} %
\def\MeanVotesFreetimeDslQuicbbr{55.82} %
\def\MeanVotesFreetimeLteQuicbbr{52.30} %
\def\MeanVotesFreetimeGggQuicbbr{47.24} %
\def\MeanVotesPlaneDaagcTcp{32.17} %
\def\MeanVotesPlaneMssTcp{30.48} %
\def\MeanVotesPlaneDaagcTcpp{34.17} %
\def\MeanVotesPlaneMssTcpp{29.32} %
\def\MeanVotesPlaneDaagcTcpbbr{32.33} %
\def\MeanVotesPlaneMssTcpbbr{28.71} %
\def\MeanVotesPlaneDaagcQuic{34.27} %
\def\MeanVotesPlaneMssQuic{33.85} %
\def\MeanVotesPlaneDaagcQuicbbr{35.03} %
\def\MeanVotesPlaneMssQuicbbr{32.88} %
\def\ABAvgDurationLab{9.58} %
\def\ABAvgDurationNet{5.40} %
\def\ABAvgDurationWork{8.04} %
\def\RateAvgDurationLab{12.06} %
\def\RateAvgDurationNet{7.01} %
\def\RateAvgDurationWork{10.00} %
\def\ABAvgFocusLab{0.00} %
\def\ABAvgFocusNet{0.50} %
\def\ABAvgFocusWork{1.08} %
\def\RateAvgFocusLab{0.03} %
\def\RateAvgFocusNet{0.57} %
\def\RateAvgFocusWork{0.93} %
\def\WorldCountries{68} %
\def\WorldTZ{0.1} %
\def\WorldCA{1.1} %
\def\WorldUS{4.5} %
\def\WorldAR{1.0} %
\def\WorldKE{3.3} %
\def\WorldDO{0.1} %
\def\WorldRU{0.8} %
\def\WorldZA{0.1} %
\def\WorldMX{0.8} %
\def\WorldBR{1.1} %
\def\WorldCO{1.0} %
\def\WorldVE{4.5} %
\def\WorldEC{0.3} %
\def\WorldPR{0.1} %
\def\WorldJM{0.3} %
\def\WorldZW{0.1} %
\def\WorldNA{0.1} %
\def\WorldBJ{0.3} %
\def\WorldGH{0.8} %
\def\WorldCI{0.1} %
\def\WorldTN{1.1} %
\def\WorldDZ{2.6} %
\def\WorldAE{0.8} %
\def\WorldIQ{0.1} %
\def\WorldVN{1.0} %
\def\WorldKR{0.3} %
\def\WorldIN{22.3} %
\def\WorldAM{0.3} %
\def\WorldUA{1.1} %
\def\WorldPL{0.7} %
\def\WorldAT{0.3} %
\def\WorldHU{0.5} %
\def\WorldRO{2.6} %
\def\WorldLT{0.3} %
\def\WorldLV{0.3} %
\def\WorldDE{0.7} %
\def\WorldBG{2.3} %
\def\WorldGR{1.4} %
\def\WorldTR{1.0} %
\def\WorldAL{1.5} %
\def\WorldHR{1.8} %
\def\WorldBE{0.1} %
\def\WorldNL{0.3} %
\def\WorldPT{1.6} %
\def\WorldES{1.0} %
\def\WorldNZ{0.3} %
\def\WorldAU{0.4} %
\def\WorldCN{0.1} %
\def\WorldTW{0.1} %
\def\WorldIT{1.2} %
\def\WorldPH{4.8} %
\def\WorldMY{3.8} %
\def\WorldSI{0.3} %
\def\WorldFI{0.3} %
\def\WorldCZ{0.3} %
\def\WorldJP{0.4} %
\def\WorldMA{2.7} %
\def\WorldUG{0.3} %
\def\WorldBA{2.3} %
\def\WorldMK{1.9} %
\def\WorldRS{9.9} %
\def\WorldME{0.1} %
\def\WorldTT{0.1} %
\def\ABAvgReplayLab{1.25} %
\def\ABAvgReplayNet{0.97} %
\def\ABAvgReplayWork{0.60} %
\def\RateAvgReplayLab{0.38} %
\def\RateAvgReplayNet{0.22} %
\def\RateAvgReplayWork{0.14} %
\def\ABAvgVideoLab{17.69} %
\def\ABAvgVideoNet{15.59} %
\def\ABAvgVideoWork{14.46} %
\def\RateAvgVideoLab{21.44} %
\def\RateAvgVideoNet{19.23} %
\def\RateAvgVideoWork{17.71} %

\title{Perceiving QUIC: Do Users Notice or Even Care?}

\author{Jan Rüth}
\orcid{0000-0002-4993-3210}
\author{Konrad Wolsing}
\author{Klaus Wehrle}
\affiliation{%
  \institution{RWTH Aachen University}
}
\email{{rueth, wolsing, wehrle}@comsys.rwth-aachen.de}

\author{Oliver Hohlfeld}
\affiliation{%
  \institution{Brandenburg University of Technology}
}
\email{oliver.hohlfeld@b-tu.de}

\renewcommand{\shortauthors}{Rüth, et al.}

\begin{abstract}
QUIC, as the foundation for HTTP/3, is becoming an Internet reality.
A plethora of studies already show that QUIC excels beyond TCP\discretionary{+}{}{+}TLS\discretionary{+}{}{+}HTTP/2.
Yet, these studies compare a highly optimized QUIC Web stack against an unoptimized TCP-based stack.
In this paper, we bring TCP up to speed to perform an eye-level comparison.
Instead of relying on technical metrics, we perform two extensive user studies to investigate QUIC's impact on the quality of experience.
First, we investigate if users can distinguish two protocol versions in a direct comparison, and we find that QUIC is indeed rated faster than TCP and even a tuned TCP.
Yet, our second study shows that this perceived performance increase does mostly not matter to the users, and they rate QUIC and TCP indistinguishable.
\end{abstract}


\maketitle

\vspace{2em}
\section{Introduction}
QUIC provides a privacy-preserving and fully-encrypted transport in user-space to ease protocol evolution.
By incorporating the ideas of several enhancements to TCP, TLS, and even HTTP, it promises an alternative Web protocol stack that is optimized for performance, security, and evolvability.
For example, it combines TLS 1.3 early-data and TCP Fast Open (TFO) to achieve 0-RTT connection establishment and data exchange.
It also offers multiplexing streams over the same connection to overcome head-of-line blocking, \eg{} by mapping HTTP/2-like streams to the transport streams.

Given QUIC's motivation to increase performance, it is no surprise that there are studies~\cite{biswal16:globecom:does,carlucci15:sac:http,cook17:icc:quic,kakhki17:imc:taking,megyesi16:icc:quick,yu17:ipccc:quic} showing that QUIC clearly outperforms the traditional Web stack (\ie{} HTTP over TLS over TCP).
Although they are subject to several limitations; \eg{} many studies utilize the Page Load Time (PLT) to measure performance.
However, it has been shown~\cite{kelton17:nsdi:webgaze,zimmermann17:inetqoe:qoepush,bocchi17:pam:web} that PLT does not correlate well with human-perceived performance.
Another shortcoming is that \emph{all} studies compare QUIC, having a highly optimized parameterization, against an off-the-shelf Web stack, which, however, offers similar optimization potential that is not used.
For example, Chromium's Google QUIC utilizes packet pacing and an initial congestion window of 32 segments.
In contrast, a stock Linux TCP stack defaults to no pacing and an initial congestion window of 10 segments, obviously disadvantaging the regular Web stack that can be avoided by parametrizing TCP similar to QUIC.

While QUIC offers new levels for protocol customization and evolution, it remains open (from a performance perspective) if switching to QUIC should be a top priority.
Hence, we ask: \emph{do humans even notice a difference, and if so, how much of a difference does it make?}
That is, does QUIC actually impact the \emph{Quality of Experience} (QoE).
We answer these questions in two user studies.

To achieve an unbiased comparison between the two stacks, we modified the Mahimahi~\cite{netravali15:atc:mahimahi} framework and incorporated the Google QUIC Web stack.
This emulation grants us full control over the network parameters, \ie{} bandwidth, delay, and queues, as well as the client and server, \ie{} enabling to modify the protocol's parameterization for comparable measurements. 
We then use video recordings of website visits in our testbed to perform studies in a controlled lab environment, using a crowdsourced online marketplace, and a voluntary crowd study to investigate the QoE of QUIC.
Our contributions are:
\vspace{-.5em}
\begin{itemize}[noitemsep,topsep=5pt,leftmargin=9pt]
\item We perform the first QoE user studies that investigate how real users perceive QUIC.
\item Our studies are unique in that they compare QUIC against a similarly parameterized TCP stack enabling eye-level comparisons.
\item We find that users \emph{do} perceive QUIC as the faster protocol in a side-by-side comparison, even against a tuned TCP stack.
\item However, in isolation, users generally do not prefer one protocol over the other if the network is sufficiently fast.
\item In slow and lossy networks, QUIC's advanced protocol design appears to cause a more satisfying loading process for our study participants, underlining QUIC's potential to improve on the long-tail of bad experiences.
\item All our study data is available at \url{https://study.netray.io}.
\end{itemize}
\vspace{-0.5em}
\afblock{Structure.}
This paper first reviews related works (\sref{sec:rw}) to see what the state-of-the-art discovered about QUIC and QoE.
Then, \sref{sec:testbed} describes our testbed and how we configured TCP similar to QUIC.
Continuing, we describe the design of our two user studies (\sref{sec:studydesign}) and evaluate the outcome of both.
Finally, \sref{sec:conclusion} concludes this paper.

\vspace{1em}
\section{Related Work}
\label{sec:rw}

\afblock{QUIC Performance.}
QUIC performance is subject to a body of studies~\cite{biswal16:globecom:does, carlucci15:sac:http, cook17:icc:quic, kakhki17:imc:taking, megyesi16:icc:quick, yu17:ipccc:quic, nepomuceno18:iscc:quic, seufert:qomex19:quic}, most compare QUIC against some combination of TCP+TLS+HTTP/1.1 or HTTP/2.

One direction of research~\cite{cook17:icc:quic, megyesi16:icc:quick} measures TCP and QUIC on publicly hosted websites---usually operated by Google.
This approach has the advantage that TCP and QUIC are \emph{likely} configured comparably on the same server.
However, these studies do not confirm any of the protocol parameterizations.
The disadvantage of these approaches is that they lack the control of the network and the server, \ie{} it becomes much harder to gain comparable measurements for QUIC and TCP since the network and the server are subject to other traffic that can have a significant impact on the measurement.
Furthermore, if the protocols are not parameterized similarly, there is no way to change this.

To enable controlled experiments, another direction~\cite{biswal16:globecom:does, carlucci15:sac:http, kakhki17:imc:taking, nepomuceno18:iscc:quic} uses self-hosted servers.
While this enables tuning the protocols to be comparable, it is not done, and the studies compare a likely unoptimized TCP against a highly tuned QUIC Web stack.
To the best of our knowledge, Yu et al.~\cite{yu17:ipccc:quic} are the only ones that investigate the impact of packet pacing in QUIC as a tuning option --- yet, they do not compare to TCP.
Similarly, Cook et al.~\cite{cook17:icc:quic} take the design of TCP and QUIC into account when looking at first and repeated connections that require less RTTs on the subsequent visit.

While self-hosting offers greater freedom than using existing sites, websites today are composed of a variety of resources that are often hosted by third parties on different servers.
Even though many studies consider websites with different resources, they often deploy only a single server~\cite{biswal16:globecom:does, carlucci15:sac:http, megyesi16:icc:quick}.
The Mahimahi framework~\cite{netravali15:atc:mahimahi} was designed to study realistic websites by replicating this multi-server nature of current websites into a testbed on which we also base our testbed (see \sref{sec:testbed}).
Nepomuceno et al.~\cite{nepomuceno18:iscc:quic} also perform a study with Mahimahi but find that QUIC is outperformed by TCP, which does not coincide with our results and related work.
We believe this is due to two reasons, first the use of the Caddy QUIC server, which is known to not (yet) perform very well~\cite{muetsch:online17:caddyvsnginx}, and second, they do not configure any bandwidth limitations.
For this reason, we chose the Google QUIC stack as it is already deployed in the Internet~\cite{rueth:pam18:quic} and likely offers good performance, which our evaluations support for the websites that we tested.

\afblock{QUIC Web Performance.}
Another line of research investigates the performance of QUIC using (visual) Web performance metrics.
Seufert et al.~\cite{seufert:icin19:quicker, seufert:qomex19:quic} investigate the QoE of YouTube video streaming by recording application layer metrics such as video quality or stalls and find no evidence for QoE improvements of QUIC over TCP.
Rajiullah et al.~\cite{rajiullah:www19:webex} achieve a similar result.
The authors conduct mobile measurements using the MONROE framework.
They evaluate technical metrics such as First Visual Change (FVC), Last Visual Change (LVC), and the RUM Speed Index (SI). 
While they find websites where QUIC has a clear impact, they conclude that overall, it has a negligible influence.

In summary, we did not find \emph{any} work investigating the \emph{user perception} of QUIC with the help of real user studies and no other work parameterized TCP stacks similar to QUIC.
In our work, we focus on both by performing a human-centered QoE evaluation of QUIC in comparison to a similarly parameterized TCP.

\section{Repeatable Protocol Performance Evaluations}
\label{sec:testbed}

\begin{table}[t]
	\centering
	\begin{tabular}{@{}l|l@{}}
		\toprule
		\ \textbf{Protocol} & \textbf{Description} \\ \midrule
		\ TCP & Stock TCP (Linux): IW10, Cubic \\ 
		\ \multirow{2}{*}{TCP+} & \unit[IW]{32}, Pacing, Cubic, \\
		\ & tuned buffers, no slow start after idle \\
		\ TCP+BBR & TCP+, but with BBRv1 as congestion control \\ 
		\ QUIC & Stock Google QUIC: \unit[IW]{32}, Pacing, Cubic \\
		\ QUIC+BBR & QUIC, but with BBRv1 as congestion control \\ \bottomrule
	\end{tabular}
	\caption{The different protocol configuration that we used in our tests.}
	\label{tab:testbed:protocols}
	\vspace{-3em}
\end{table}

To compare the performance of website load processes subject to different protocol configuration, we reuse our Mahimahi-based testbed from~\cite{wolsing:anrw19:quicperf}.
In that work, we analyzed QUIC's performance subject to several Web performance metrics such as FVC, LVC, SI, Visual Completeness 85\% (VC85), and PLT.
To this end, we embedded the gQUIC Web stack as well as an NGINX Web server in Mahimahi and repeatedly requested a set of previously recorded websites using the Chromium browser, which we automated using Browsertime~\cite{browsertime}.
In the following, we highlight the key properties of the testbed as well as the configurations that we applied for our QoE user studies.

\afblock{Protocol Parametrization.}
By default, Mahimahi does not alter the Linux stack's TCP configuration and uses the shipped defaults.
Please note that we refer to the \emph{configurations and parameterizations} and not each implementation's bare performance when we refer to the stacks as tuned.\footnote{We believe that the Linux TCP stack, given its age, has a more efficient code-base.}
We extend Mahimahi in that it is simple to reconfigure the stacks before requesting websites.
\tab{tab:testbed:protocols} summarizes the different protocol configurations that we apply to QUIC and TCP.
To this end, our \emph{TCP+} is configured to match gQUIC's default of an initial congestion window of 32 segments, using pacing (with Linux's defaults of an initial quantum of ten and a refill quantum of two segments) and not falling back to the initial window after idle.
Further, we enlarge the send and receive buffers according to the bandwidth-delay product (BDP) of the underlying network to allow full utilization.
Moreover, since we already found a significant impact of the congestion control in~\cite{wolsing:anrw19:quicperf}, we further use a variant of TCP and QUIC that utilized BBRv1.\footnote{BBRv2 was not yet available at the time of testing.}

A website visit in our testbed is done using a fresh Chromium browser with an empty cache.
Starting from scratch has several implications on the protocols: it means that QUIC will \emph{not} perform a 0-RTT connection establishment but a 1-RTT handshake.
Since there is no support for TLS 1.3 early-data in Chromium (as of June 2019) and a limited and challenging deployment of TFO in the Internet, this still gives a 1-RTT advantage of QUIC over TCP+TLS+HTTP/2.
In~\cite{wolsing:anrw19:quicperf}, we found that this 1-RTT advantage is the primary factor for QUIC outperforming the traditional Web stack in non-lossy environments when looking at technical metrics.
QUIC, similar to early-data and TFO, suffers from replay attacks, which have an especially large surface in distributed clusters~\cite{fischlin:eurosp17:0rttreplay} when requests are non-idempotent.
To this end, \eg{} Cloudflare only allows non-parameterized GET requests via TLS 1.3 0-RTT~\cite{cloudflare:online17:intro0rtt}.
In summary, there is currently a lack of signaling idempotency through the stack such that 0-RTT could be easily enabled on a large scale.
Thus, we believe that comparing a 1-RTT QUIC with a 2-RTT TCP/TLS reflects a likely Internet-wide deployment of both, even though we will see 0-RTT QUIC and TLS early-data in parts.

\afblock{Websites Selection.}
Selecting websites that are representative of large parts of the Web is challenging.
We base our selection of websites on~\cite{wijnants2018http}, which ultimately derived a list of 40 sites from the Alexa Top 50 and Moz Top 50.
They were chosen to have a high variation in size (number of objects and their sizes) as well as contacted IP addresses (multi-server nature).
A visualization of this variation (in terms of multi-server nature and website size distribution) is given in the original paper~\cite{wijnants2018http} as well as our previous work~\cite{wolsing:anrw19:quicperf}.
For this paper, we restrict our view to 36 of the 40 websites.
We had to exclude four sites since we were unable to replay them accurately, and one was a private project site.

\begin{figure}[t]
\centering
\includegraphics[width=\columnwidth]{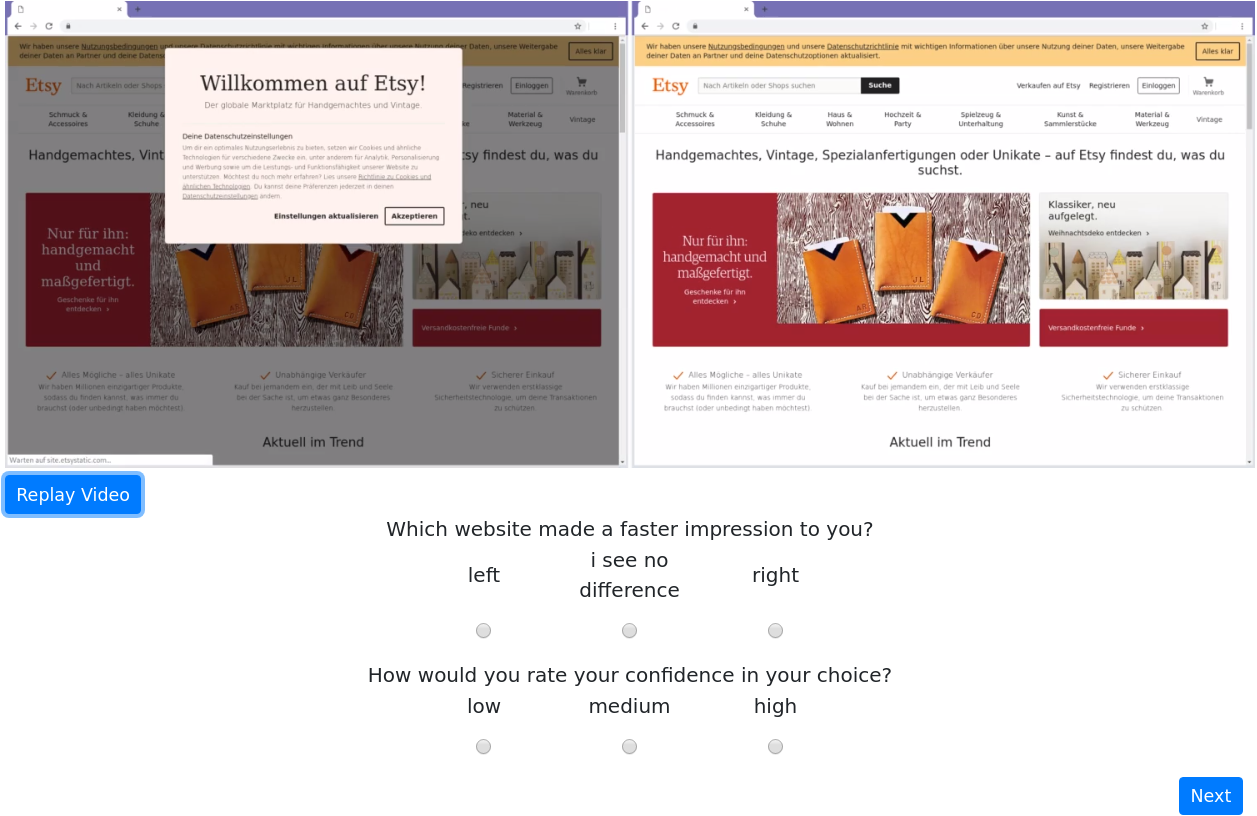}
\caption{Example screenshot from the A/B study with video and questions. Note that usually the questions are hidden and pop up one after the other.}
\label{fig:abstudy-screenshot}
\vspace{-1em}
\end{figure}

\begin{table}
	\centering
	\begin{tabular}{l|r|r|r|r} 
		\toprule
		\textbf{Network} & \textbf{Uplink} & \textbf{Downlink} & \textbf{min.\ RTT} & \textbf{Loss} \\ \midrule
		DSL		&    \unit[5]{Mbps} &   \unit[25]{Mbps} &  \unit[24]{ms} &  \unit[0.0]{\%} \\ 
		LTE		&  \unit[2.8]{Mbps} & \unit[10.5]{Mbps} &  \unit[74]{ms} &  \unit[0.0]{\%} \\ 
		DA2GC	& \unit[.468]{Mbps} & \unit[.468]{Mbps} & \unit[262]{ms} &  \unit[3.3]{\%} \\
		MSS		& \unit[1.89]{Mbps} & \unit[1.89]{Mbps} & \unit[760]{ms}&  \unit[6.0]{\%} \\  \bottomrule
	\end{tabular}
	\caption{Network configurations. Queue size is set to \unit[200]{ms} except for DSL with \unit[12]{ms}.}
	\label{tab:testbed:network}
	\vspace{-2.5em}
\end{table}

\afblock{Network Parameter Selection.}
Within our user study, we want to cover the user-perception in ``good'' networks as well as in ``bad'' networks.
To this end, we select four different network settings, which we summarize in \tab{tab:testbed:network}.
We chose the median bandwidth of German households according to the federal network agency~\cite{zafaco:online18:breitbandmessung}, which we refer to as DSL.
This network has no artificial random loss, and we set a low minimum round-trip time (RTT) to which the queue adds further jitter (up to \unit[12]{ms}).
Similarly, we use median bandwidth values for German mobile users, which we refer to as LTE.
Even though it is a wireless link, we set no artificial loss as the link-layer would recover it.
Still, the min.\ RTT is already higher.
Furthermore, we allow up to \unit[200]{ms} of queuing.
Lastly, we take network parameters for two ``bad'' in-flight WiFi networks that connect either via LTE to the ground (DA2GC) or via an in-flight satellite connection (MSS).
Those parameters have been established in~\cite{rula2018mile} and are characterized by low bandwidths and high delays as well as high random loss.

\afblock{Producing Videos.}
To record videos to show in our user study, we automate the Chrome browser to visit the 36 websites replayed in our testbed at least 31 times while recording the browser window.
This enables us to derive technical and visual metrics (FVC, LVC, PLT, SI, VC85) with a certain accuracy (for later comparison) and subsequently select a video that closely fits a ``typical'' recording by taking the video that is closest to the average PLT inspired by~\cite{zimmermann17:inetqoe:qoepush}.

\section{User Perception of Protocol Speed}
\label{sec:studydesign}

The key focus of our study is to evaluate the effect of protocol performance on user perception, tested in two studies.

\afblock{Study 1 (A/B): Do Users Notice?}
We begin by performing a just noticeable difference test to identify if users notice a protocol switch.
The study design involves a pairwise comparison where two recordings of the loading processes of the same website with different protocol configurations but the same network setting are shown side-by-side (rendered into a single video) to participants (see \fig{fig:abstudy-screenshot}).
This pairwise comparison allows us to detect even subtle differences in the loading processes.
After watching the video, the participants are asked to answer if the left or right video was the faster one or if they cannot decide.
We furthermore ask them to rate their confidence in their choice.

\begin{figure}[t]
\centering
\includegraphics[width=0.662\columnwidth]{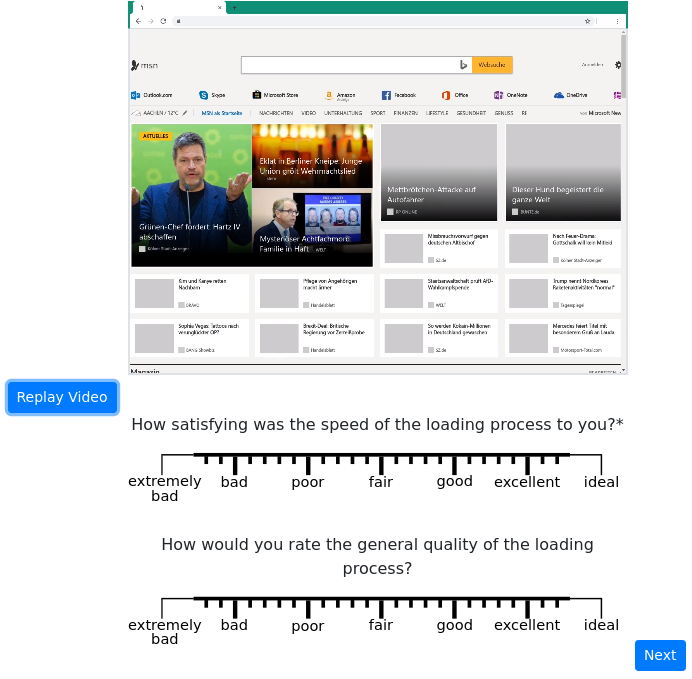}
\caption{Example screenshot from the rating study with video and questions. Note that usually the questions are hidden and pop up one after the other.}
\label{fig:ratestudy-screenshot}
\vspace{-2em}
\end{figure}

\afblock{Study 2 (Rating): Do Users Care?}
While the first study informs us if protocol switches lead to perceivable differences, it does not tell how users rate the perceived quality of the loading process.
We answer this aspect in a second study, in which we only present one video (see \fig{fig:ratestudy-screenshot}) to let participants rate \emph{i)} their satisfaction with the loading speed and \emph{ii)} the general quality of the loading process.
Both ratings are obtained on a seven-point linear scale~\cite{itu:recommendataion03:qoemetric} ranging from extremely bad over bad, poor, fair, good, excellent to ideal, mapped to values from 10 to 70 with equidistance selectable by participants with a granularity of 1.
To set a context for assessing speed perception, we ask the participants consider being in a particular environment for this study: imaging being \emph{i)} at work, \emph{ii)} in their free time, or \emph{iii)} on a plane.

We implemented the user studies using TheFragebogen~\cite{guse:qomex19:thefragebogen} and host it on our infrastructure.
Each study begins with a tutorial that explains the purpose and the procedure of the study.
By informing the participants on the study goals and its procedure, it also aims at reducing noise in the responses.

\afblock{Pilot Study.}
We tested our study in a pilot study before releasing it.
It involves volunteers (friends and colleagues) testing our system to see if people that are unfamiliar with the study can perform it.
The results of the pilot study are not used for evaluation, and participants did not participate in both the pilot and the later study to limit bias by training effects.
The main feedback was that people were overwhelmed when the videos start without a tutorial up front, which we added.
Secondly, we also rendered a Web browser window around each video (also shown in the figures) as we got the feedback that otherwise, people were unsure about the bounds of each website.

\subsection{Performing the User Studies}

We utilize three different subject groups for our user study.

\afblock{Lab Study.}
First, we perform a controlled lab study with both the rating and the A/B study (the beginning is randomized), where we can monitor and supervise the study.
Since the lab supervisor monitors the user behavior during the lab study (\eg{} to check that participants properly conduct the test, \ie{} actually watching the videos and not clicking randomly on the scales), it serves as our control data to evaluate the other two uncontrolled crowdsourcing groups.
As this control group is rather small, we only consider five domains (wikipedia.org, gov.uk, etsy.com, demorgen.be, nytimes.com), which are diverse in website size such that the overall duration for each participant is roughly \unit[10]{min}.
This constraint leads us to show 28 videos for the A/B study and 11 in the work, 11 in the free time and only five, due to the increased video length, in the plane environment for the rate study.

\afblock{Crowdsourcing Studies.}
We employ crowdsourcing to enlist a larger number of participants.
As a second group, we recruit paid Microworkers~\cite{microworker:online19:landingpage} participants. 
We follow the platform's guidance and offer \unit[0.75]{USD} for a study between 10 and \unit[15]{min} length and allow participation of a user in each study only once.
After a Microworker (µWorker) completed the study, she can redeem her payout using a code that we display at the end of the study.
We show 26 videos in the A/B study and again 27 (11 work, 11 free time, and 5 plane) in the rating study.
Third, we advertise the studies on social media to recruit regular Internet users.
As we expect unwillingness to perform a lengthy study, we show only 14 videos (A/B) and 15 videos (6 work, 6 free time, 3 plane) in the rating study.

\afblock{Conformance Filtering.}
While the controlled lab study helps us in judging the quality of the crowd-sourced results, we take extensive measures to ensure valid results since we suspect that at least on the Microworker platform, people will cheat the system to solve the study quickly.
To this end, we implement 7 rules that are used to filter invalid results:
\textbf{R1:} A video in the study has not been played. 
\textbf{R2:} A video has stalled. 
\textbf{R3:} There is a focus loss event (\eg{} website not the active tab or window not in the foreground) for longer than \unit[10]{sec} during the study.
\textbf{R4:} A vote was placed before the FVC.
\textbf{R5:} A study took longer than \unit[25]{min} or a question took longer than \unit[2]{min}.
\textbf{R6:} A randomly placed control video was answered wrong.
In the A/B study, we embed significantly delayed variants of the left or right video or have the same video on both sides.\footnote{Since even in the lab study people claimed to see a difference, we allowed left or right as a valid answer if the confidence was low.}
In the rating study, we embed a very quickly rendering website and a very slow one; we expect the rating to be at least 10 points apart.
\textbf{R7:} A control question was answered wrong.
Every fifth video includes an additional question that asks for the color of the browser frame, which is still visible while answering the question (see, \eg{} \fig{fig:ratestudy-screenshot} having a green browser frame).
Each video is assigned a random color from red, green, and blue; we chose the exact colors to be colorblind safe.

\begin{table}
	\centering
	\setlength{\tabcolsep}{0.25em} %
	\begin{tabular}{rl|c|c|c|c|c|c|c|c} 
		\toprule
		 & &\textbf{-} & \textbf{R1} & \textbf{R2}  & \textbf{R3} & \textbf{R4} & \textbf{R5}   & \textbf{R6} & \textbf{R7} \\ \midrule
		 \multirow{2}{*}{Lab} & A/B                &     - &      - &      - &      - &     - &      - &      - &  \underline{\ABFilterLabvii} \\ 
	                                         & Rating            &     - &      - &      - &      - &     - &      - &      - &  \underline{\RateFilterLabvii} \\ \midrule
		 \multirow{2}{*}{µWorker}  & A/B         & \ABFilterWork & \ABFilterWorki & \ABFilterWorkii & \ABFilterWorkiii & \ABFilterWorkiv & \ABFilterWorkv & \ABFilterWorkvi & \underline{\ABFilterWorkvii} \\
							& Rating    & \RateFilterWork & \RateFilterWorki & \RateFilterWorkii & \RateFilterWorkiii & \RateFilterWorkiv & \RateFilterWorkv & \RateFilterWorkvi & \underline{\RateFilterWorkvii} \\ \midrule
		 \multirow{2}{*}{Internet}  & A/B   	& \ABFilterNet & \ABFilterNeti & \ABFilterNetii & \ABFilterNetiii & \ABFilterNetiv & \ABFilterNetv & \ABFilterNetvi & \underline{\ABFilterNetvii} \\ 
						       & Rating	& \RateFilterNet & \RateFilterNeti & \RateFilterNetii & \RateFilterNetiii & \RateFilterNetiv & \RateFilterNetv & \RateFilterNetvi & \underline{\RateFilterNetvii} \\ \bottomrule
	\end{tabular}
	\caption{Participation in our studies and results after each filter rule, final participation are underlined.}
	\label{tab:study:participants}
	\vspace{-2em}
\end{table}

\tab{tab:study:participants} summarizes the participation and how many results we removed due to each of the filters.
Since we allow each µWorker to only participate once\footnote{We cannot filters users with multiple µWorker accounts.} in each study, and we suspect Internet users not to repeat the studies, these numbers should be close to the true number of individuals participating.
Focus loss (R3) and voting before the FVC (R4) filtered the most results.

\afblock{Ethical Considerations.}
Our study design follows standard guidelines for conducting crowdsourcing QoE studies~\cite{HossiBestPractices}.
Each participation (in the lab, from the Internet or as paid workers) takes place voluntarily.
For the Microworker platform, we follow the platform's guidelines for payment.
We chose not to pay lab participants to allow participation without monetary pressure.
The user studies clearly state which data we are gathering, and only after completion of the study, this data is uploaded securely to our servers.
Regarding the stimulus, the content in all videos does not show any sensitive material, \eg{} violence, abuse, or other questionable content. 
In case of difficulties or errors, we are reachable via email on the study website and directly via the Microworker platform.

\subsection{Study Agreement}
We first compare our controlled lab study against both crowdsourced studies.
For the A/B study, on average, lab participants took \unit[\ABAvgVideoLab]{s}, µWorker \unit[\ABAvgVideoWork]{s}, and Internet user \unit[\ABAvgVideoNet]{s} per video.
The rating study took a little more time, lab participants took \unit[\RateAvgVideoLab]{s}, µWorker \unit{\RateAvgVideoWork}{s}, and Internet users \unit[\RateAvgVideoNet]{s} per video.
We found lab participants replay videos more often, especially in the A/B study.
Regardless of group, faster networks resulted in more replays, which might already indicate that it is harder to spot differences.
Regarding demographics, 76\% to 79\% were male.
Within the Internet and the Lab group, the majority was younger than 24 years, for the µWorkers two-thirds were between the age of 25 to 44 years. 

\begin{figure}[t]
\centering
\includegraphics[width=\columnwidth]{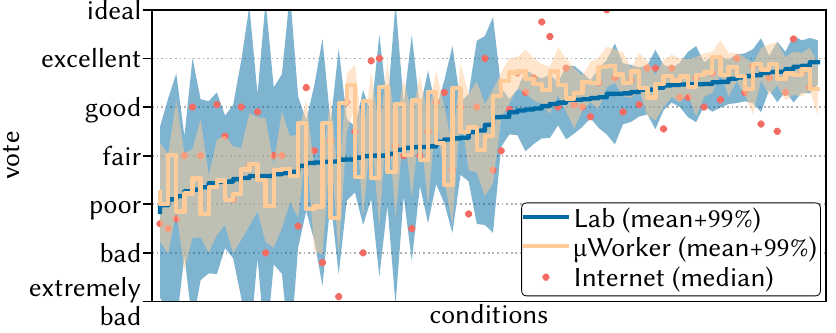}
\vspace{-1em}
\caption{Rating study votes over all lab-tested conditions (ordered by mean vote of the lab participants). We show the 99\% confidence interval. Internet values are not normally distributed, and thus, we show the median.}
\label{fig:study:rate-vote_cmp}
\vspace{-1.5em}
\end{figure}

\fig{fig:study:rate-vote_cmp} shows the agreement of all three groups in the rating study across all conditions on the x-axis (conditions in the rating study are single video and video pairs in the A/B study).
The lab, as well as the µWorker data, is normally distributed.
Thus, we show the mean and the 99\% confidence interval (shared area) of the votes.
We find that the µWorkers seem to fall mostly within the confidence intervals of the lab study, and hence, we believe that these votes are legit.
In contrast, for Internet votes, we are unable to estimate the distribution and thus show the median of the votes.
As is visible from the figure, the Internet group deviates most from the other two, and the number of votes that we were able to collect is lower.
Consequently, we exclude it from further discussions, highlighting the challenge when trying to collect voluntary user data while setting a high standard on compliance with basic rules.

For the A/B study, in general, the agreement follows a similar scheme (not shown), but outliers look more severe due to the 3-point-scale (left, right, no diff.).
We manually inspect the significant outliers in both studies, and we found that the websites are structurally very different.
\fig{fig:abstudy-screenshot} shows such a case, after loading the actual content, a welcome banner pops up.
Participants in the lab study made their decision after the banner loaded (see left video) while people in the crowdsourced data seem to vote earlier according to when the actual website content is shown, and those decisions often do not agree across different protocol versions.

\subsection{Do Users Notice a Difference?}
\begin{figure}[t]
\centering
\includegraphics[width=\columnwidth]{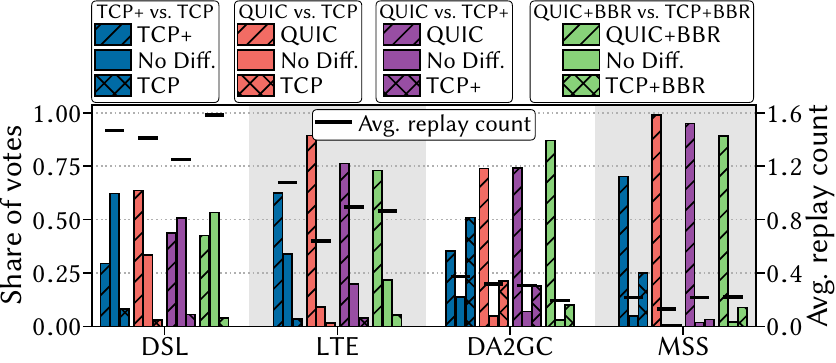}
\vspace{-1em}
\caption{A/B study mean votes for each protocol combination depending on the network configuration.}
\label{fig:study:ab-meanvote}
\vspace{-1em}
\end{figure}

To answer the questions if users actually notice a difference between the different protocol variants, we look at the A/B study and compare their votes.
\fig{fig:study:ab-meanvote} shows the share of votes for preferring a specific protocol variant in the four different network settings across all websites.
The colors denote different pairs of protocols under comparison, the hatches signal preference for one or the other.
Furthermore, we display the average replay count as vertical lines for each group of comparisons.
So, \eg{} in the DSL setting, slightly over 25\% prefer TCP+, over 60\% see no difference, and less than 10\% prefer TCP, but on average, people replayed the video roughly 1.4 times.
In general, we observe that the agreement for observing a difference rises when the networks become slower.
For example, in the DSL setting, for all but the QUIC vs.\ TCP comparison (red), most participants do not see a difference.
The comparably high average replay count expresses the difficulty of spotting a difference in the DSL network.
Still, in general, more people experience the protocol variant quicker that is supposed to be faster.
Lowering the bandwidth towards the LTE setting, the majority of participants now clearly vote the supposedly better variant (this confidence is also backed by the lower replay counts).
In the slower networks, there are slight differences.
For DA2GC, TCP is now favored in contrast to our tuned variant (TCP+), we always found more retransmissions for TCP+ (on avg. $\times$1.5 but up to $\times$4.8) which may be explained by the comparably high initial congestion window leading to early losses.
In contrast, QUIC seems to not suffer from the same problems (even though similarly configured to TCP+) as our participants experience it faster, we suspect that QUIC's large SACK ranges enable it to progress further and that the independent stream processing allows earlier renderings of the page.
Looking at the MSS network, the observation from DA2GC is now again reverted for TCP vs.\ TCP+; the increased bandwidth reassures our earlier assumption.
For the other protocol variants, the picture from DA2GC is now even stronger with even more votes towards the supposedly faster variants.
Again, the higher random loss-rate in this network backs our previous impression.

\takeaway{In general, people do see a difference and indeed perceive QUIC as the faster protocol over TCP and even over a tuned TCP variant. However, in networks with high bandwidths, perceiving a difference seems to be more challenging.}

\subsection{Do Users Even Care?}

\begin{figure}[t]
\centering
\includegraphics[width=\columnwidth]{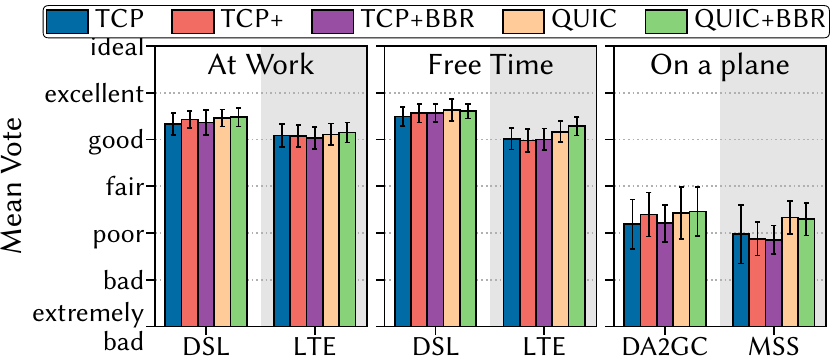}
\vspace{-1em}
\caption{Rating study votes per protocol choice in the different settings for the µWorker group. Error bars show 99\% confidence intervals.}
\label{fig:study:rate-votesoverall}
\vspace{-1em}
\end{figure}

We continue to answer the second question, whether users care, or more specifically, we want to investigate if users perceive the already uncovered speedups as actually increasing the performance or if they cannot tell in isolation.
To this end, we look at the results of the rating study, which we overview in \fig{fig:study:rate-votesoverall}.
In general, we observe that the work and free time scenario are rated similarly with a slight tendency towards better scores in the free time setting for DSL.
In contrast is the plane setting (only having videos using the emulated in-flight networks) that shows only poor results.

When looking at the results within a network setting, we see only little variance between the different protocols, and the confidence intervals mostly overlap.
When we test the different distributions for significance (using a significance level of 99\% and an ANOVA test), we do not find any significant protocol/network configuration that the users rate better.
When we lower the confidence level to 90\%, three settings are significantly different.
First, in the LTE free time setting, QUIC+BBR is rated statistically more satisfying than TCP+BBR.
BBR again makes the difference in the plane environment and the MSS network.
In the same setting, also QUIC without BBR is generally rated faster than TCP+.
Thus, in general, there is little difference between the protocol variants.
We now take a look at the specific websites where changes matter.

\afblock{Where it Makes a Difference.}
In the DSL setting, eight websites show significant differences.
Four of them rate QUIC faster than TCP, one faster than TCP+, and three rate QUIC+BBR faster than TCP+BBR.
Spotify.com shows the largest difference with BBR. 
The website is small, but the browser has to contract many hosts.
Still, we find small and also large websites that profit from QUIC.

In the LTE setting, only five pages show a significant difference.
Our participants favor TCP+ over TCP once, otherwise QUIC over TCP+ (twice) and TCP (once).
When using BBR, QUIC is favored once over TCP+.
Regardless of congestion control, QUIC is roughly rated 10 points better, \ie{} a whole quality level.
Again, the websites show a wide variety of sizes and contacted hosts.

For DA2GC, we again find only five websites with a significant difference.
Apache.org, a relatively small website in terms of size and resources, is preferred when delivered via QUIC in contrast to TCP and TCP+.
When using BBR, google.com, gov.uk, and nature.com are perceived faster using QUIC.
Lastly, w3.org is rated over 15 points faster when using QUIC in contrast to TCP+.

For MSS, we find three pages.
Wordpress.com is favored in the QUIC variant over both TCP variants, a website with few resources, small in size, and less than ten contacted hosts.
Gravatar.com on TCP is liked less.
Apache.org is favored when BBR is used.

\afblock{Correlation to Technical Metrics.}
We next investigate which technical metrics (FVC, SI, VC85, LVC, and PLT) best reflect our participants' ratings.
To do so, we calculate the Pearson's correlation coefficient of the votes compared to the technical metrics by first calculating the mean vote for each website and combining it with the technical metric.
We chose Pearson's (\eg{} over Spearman) because we are interested to see how well the linearity of the metric reflects the users' choices.
Thus, we would assume a high negative coefficient when high vote scores are correlated with low metric scores (\ie{} a quick loading) and vice versa.

\begin{figure}[t]
\centering
\includegraphics{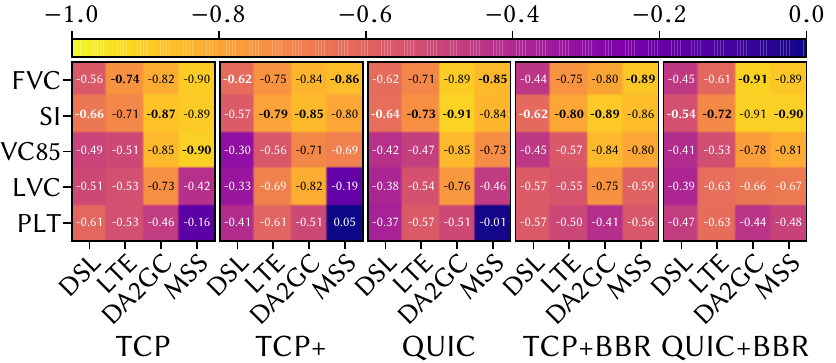}
\vspace{-1em}
\caption{Pearson's correlation coefficient heatmap for different technical metrics to the user ratings in the different settings. High negative values (-1.0) are desired and mean that technical metric and user rating correlate perfectly. Largest correlation are in bold. For DSL/LTE, we chose the votes from free time setting.}
\label{fig:study:rating_corr}
\vspace{-1em}
\end{figure}

\fig{fig:study:rating_corr} shows a heatmap for the different correlation coefficients in the different network settings subject to the different protocols.
As the figure shows, in general, SI shows the largest correlation even though we find that in speedy networks such as DSL, the correlation goes down, and for slower networks, it goes up.
However, this is not limited to SI but also holds for the other metrics, which seems to be in line with earlier findings that showed larger confidence in these networks.
Opposing SI, we find PLT to have the worst correlation to our users' ratings, thus reinforcing related works.

\takeaway{In general, our participants did not care about the protocol when giving no direct comparison even though, in some cases, QUIC showed a small tendency to be preferred. 
When looking at individual websites, their size in terms of bytes and objects was not an obvious reason to prefer QUIC; only many contacted systems seem to point towards QUIC.
In the challenged networks, we would have suspected that big websites would make a greater difference, but we found the opposite, which is likely due to the overall long loading times, where only small websites may show a perceivable deviation.
Furthermore, we found that the Speed Index shows the highest correlation with our participants' votes.}

\section{Conclusion}
\label{sec:conclusion}
Motivated by related works that compare a highly-tuned QUIC Web stack against unoptimized TCP-based Web stacks, we performed two Quality of Experience studies to compare a state-of-the-art QUIC stack against tuned versions of TCP-based stacks.
Our first study confirms related works that claim that QUIC indeed outperforms TCP-based stacks.
Our user study shows that actual users can \emph{perceive} these often small technical performance differences.
Thus, users rate QUIC as faster.
However, if network speeds increase, the difficulty of spotting a difference rises.
In a second study, we investigate if people actually prefer the QUIC-delivered version of a website over TCP by showing them recordings of loading processes in isolation.
Our results now indicate that the differences seem to become negligible, and users do not perceive one protocol as significantly faster; only in slower networks, people seem to tend towards preferring QUIC.
Furthermore, our results support the use of the Speed Index as we found it to best correlate to user perception for the tested protocols and websites.
In summary, our results show that QUIC, while radically evolving the transport layer, will \emph{not} tremendously increase user satisfaction by just enabling it.
While QUIC \emph{is the future to evolve and enhance the transport layer}, switching today should not be purely motivated by performance, the enhanced privacy and its future proof design are what sets it apart and may give improved performance in the future.

\begin{acks}
This work has been funded by the DFG as part of the CRC 1053 MAKI subprojects B1 and A2, and the SPP 1914 REFLEXES.
\end{acks}
\bibliographystyle{ACM-Reference-Format}
\balance
\bibliography{paper}

\end{document}